# DETERMINACIÓN DE LA SOLUBILIDAD DEL DODECANO EMPLEANDO LA TEORÍA LIFSHITZ-SLYOSOV-WAGNER

Loredana Espinoza, Kareem Rahn-Chique*, Issarly Rivas, German Urbina-Villalba

Instituto Venezolano de Investigaciones Científicas (IVIC), Centro de Estudios Interdisciplinarios de la Física (CEIF), Carretera Panamericana Km. 11, Aptdo. 20632, Caracas, Venezuela. Email: krahn@ivic.gob.ve

**Resumen**  Se determinó la solubilidad del dodecano empleando la teoría de Lifshitz-Slyosov-Wagner para el fenómeno de maduración de Ostwald. A tal objeto se prepararon nanoemulsiones de dodecano en agua estabilizadas con dodecilsulfato de sodio y se evaluó la variación del radio cúbico promedio de sus gotas en función del tiempo a diferentes fracciones de volumen ($\phi = 10^{-3} - 10^{-5}$). Se encontró que el sistema más diluido ($\phi = 1 \times 10^{-5}$) arrojó el valor de solubilidad más confiable ($5{,}5 \times 10^{-9}$ m$^3$/m$^3$).

**Palabras Clave**  Emulsion, Dodecano, Solubilidad, Ostwald, LSW, Nano

## 1. INTRODUCCIÓN

La solubilidad acuosa de un compuesto es la cantidad máxima del mismo que puede disolverse en una cantidad fija de agua a una cierta temperatura. En el caso de sustancias líquidas inmiscibles como los alcanos, la cantidad disuelta es imperceptible al ojo humano, y muy difícil de determinar mediante técnicas espectrofotométricas basadas en la Ley de Beer-Lambert, dado que este tipo de sustancias no absorben de manera sensible en el espectro UV-visible, o en su defecto, poseen picos de absorción en la misma región del agua. En estos casos, el uso de la teoría Lifshitz-Slyosov-Wagner (LSW) aplicada al fenómeno de maduración de Ostwald, permite determinar la solubilidad del aceite partiendo de la variación del tamaño promedio de gota de una emulsión aceite/agua.

Se describe con el término de Maduración de Ostwald (Ostwald Ripening) al fenómeno mediante el cual gotas de diferente tamaño intercambian aceite a distancia por medio de difusión molecular a través del solvente. Este mecanismo de desestabilización tiene su origen en la dependencia inversa de la presión interna de las gotas con su radio (ecuación de Laplace). De allí que el potencial químico de las moléculas que constituyen una gota pequeña sea mayor que el de aquellas que constituyen una gota grande. El sistema tiende al equilibrio solubilizando moléculas de las gotas pequeñas. Como resultado, la concentración de moléculas de soluto en una vecindad de las gotas (aproximadamente igual a su radio) es distinta a aquella en el seno del solvente. Esto se expresa analíticamente mediante la ecuación de Kelvin [Thomson, 1871]:

$$S(r) = S(r = \infty) \exp(2\gamma V_m / r R_g T) = S(\infty) \exp(\alpha/r) \quad (1)$$

Donde $S(r)$ es la solubilidad de una gota de aceite de radio r (en m$^3$/m$^3$), $S(r = \infty)$ es la solubilidad de una fase macroscópica (sobrenadante) de aceite en agua, es decir, cuando el aceite forma una interfase plana ($r = \infty$) de manera que no existe diferencia de presión entre él y el agua, $\gamma$ es la tensión interfacial aceite/agua, $V_m$ el volumen molar del aceite (m$^3$/mol), $R_g$ la constante universal de los gases, T la temperatura, y $\alpha$ el "radio capilar del aceite".

La dependencia de la solubilidad con el radio de la gota conlleva a una situación compleja en presencia de un conjunto de gotas de diferente tamaño. Tal distribución es típica de una emulsión, y en general de cualquier dispersión coloidal. La variación temporal de una distribución de tamaños de gotas en el límite de dilución infinita es descrita por la teoría de LSW propuesta simultáneamente por Lifshitz y Slesov (Slyosov) y separadamente por Wagner en 1961 [Lifshitz, 1961; Wagner, 1961]. La teoría parte de las siguientes premisas:

a) Las partículas están fijas en el espacio
b) El sistema está infinitamente diluido por lo que no existen colisiones entre gotas
c) La concentración de fase interna es la misma a través de toda la fase externa excepto en la cercanía de las partículas, y varía en su vecindad de acuerdo a la ecuación de Kelvin.

d) Las moléculas de fase interna se mueven por difusión entre partículas.

Como resultado del proceso existe un radio crítico de la dispersión que crece con el tiempo. A tiempos largos, en el llamado "régimen estacionario" la variación del radio cúbico promedio puede estimarse mediante una ecuación simple:

$$\frac{dr_c^3}{dt} = 8\, S(\infty) \gamma V_m D / 9 R_g T \qquad (2)$$

Donde $D$ el coeficiente de difusión (m²/s) de las moléculas de aceite (disueltas de la fase acuosa) y $r_c$ el radio crítico de la dispersión, el cual es equivalente al radio promedio (en número) de sus gotas. Gotas con tamaños inferiores al promedio se disuelven, y gotas con tamaños superiores al promedio crecen. Como $r_c$ evoluciona en el tiempo, la identidad de las gotas que crecen y se disuelven también cambia. El resultado a tiempos largos es una distribución de tamaño de gotas característica, similar a una log-normal pero con cola hacia la izquierda (hacia menores tamaños), que también recibe el nombre de LSW.

La variación del radio cúbico respecto al tiempo se utiliza comúnmente a fin de caracterizar la velocidad de ocurrencia del fenómeno de maduración [Binks, 1998; Kabalnov, 1990a; Taylor, 1998; Kabalnov, 2001]. Sin embargo, de acuerdo a nuestras simulaciones y a la evidencia experimental reciente [Nazarzadeh, 2012], el tamaño de gota sólo crece por maduración de Ostwald cuando las gotas más pequeñas se disuelven completamente [Urbina-Villalba, 2014], de otra manera el radio promedio disminuye por intercambio molecular. Por ello es muy factible que el incremento del tamaño de gota promedio producto de los fenómenos de floculación y coalescencia se asocie erróneamente a maduración de Ostwald. De hecho, la variación de la concentración de gotas con el tiempo, sigue una cinética de segundo orden similar a la del fenómeno de agregación.

Si se garantiza que la completa disolución de las gotas pueda darse evitando el fenómeno de floculación, es posible calcular la solubilidad de un compuesto líquido inmiscible con agua midiendo la variación del radio promedio de gota de su emulsión con respecto al tiempo [Kabalnov, 1990b]. De acuerdo a la ecuación (2):

$$S(\infty) = \left(\frac{dr_c^3}{dt}\right) 9 R_g T / 8 \gamma V_m D \qquad (3)$$

Se conoce que las emulsiones están sometidas a varios procesos de desestabilización como floculación, coalescencia, formación de crema y maduración de Ostwald. Sin embargo, es posible utilizar la ecuación (3) si se diluye el sistema lo suficiente y se sintetizan gotas muy pequeñas a fin de evitar el proceso gravitatorio. Las gotas de tamaño nanométrico no son susceptibles al efecto gravitatorio en menos de 6 meses a menos que se agreguen. Por otra parte, la agregación puede retrasarse haciendo emulsiones poco concentradas y utilizando surfactantes que creen barreras repulsivas entre las gotas, de forma que los encuentros entre ellas disminuyan, y se evite la floculación irreversible.

En este trabajo (proyecto científico de la Br. Espinoza) se determinó la solubilidad del dodecano empleando la ecuación (3). Para esto, se prepararon dispersiones de dodecano en agua a diferentes fracciones de volumen, y se determinó la variación del tamaño promedio de las mismas en función del tiempo.

## 2. PROCEDIMIENTO EXPERIMENTAL

### 2.1. Materiales

El n-dodecano (Merck, 98% de pureza) fue purificado eluyéndolo a través de una columna de alúmina. El resto de los reactivos: dodecilsulfato de sodio (SDS, Merck), cloruro de sodio (NaCl, Merck, 99.5% de pureza) e isopentanol (iP, Scharlau Chemie, 99% de pureza) fueron empleados tal y como fueron recibidos. El agua utilizada en todos los experimentos fue destilada y desionizada (conductividad < 1 $\mu S cm^{-1}$ a 25 ºC) usando un purificador Simplicity (Millipore, USA).

### 2.2. Síntesis de nanoemulsiones

Se preparó una nanoemulsión madre a partir de la dilución de un sistema en equilibrio compuesto por tres fases, agua (W) + cristal líquido (CL) + aceite (O) [Rahn-Chique, 2012] de composición: 10 % p/p SDS, 8 % p/p NaCl, 6,5 % p/p iP y una fracción en peso de agua igual a: $f_w$ = 0,20 (equivalente a una fración de volumen de $\phi$ = 0.84). Esta mezcla fue diluida (a 5 % p/p SDS, 3,3 % p/p iP, $\phi$ = 0,44) añadiéndola sobre agua bajo agitación constante (1000 r.p.m.) hasta obtener una nanoemulsión madre (NM) con un radio promedio de gota de r = 76,3 nm. Seguidamente, ésta emulsión fue diluida a diferentes fracciones de volumen (1 x $10^{-3}$, 5 x $10^{-4}$, 1 x $10^{-4}$, 5 x $10^{-5}$ y 1 x $10^{-5}$) empleando soluciones acuosas de SDS. Todos los experimentos fueron realizados a T = 298 K, manteniendo una concentración final de SDS de 8 mM.

### 2. 3. Evolución de radio cúbico en función de tiempo

El tamaño de gota fue determinado utilizando un Goniómetro BI-200SM (Brookhaven, USA). Las mediciones se prolongaron por un espacio de 7 semanas.

### 2.4 Parámetros necesarios

A fin de calcular la solubilidad del dodecano es necesario conocer $V_m$ (2,27 x 10$^{-4}$ m$^3$/mol), $D$ (5,40 x 10$^{-10}$ m$^2$/s) y $\gamma$ (9,16 x 10$^{-3}$ N/m). El valor de la constante de difusión fue obtenido de la bibliografía [Sakai, 2002] y típicamente es del orden de 10$^{-10}$ m$^2$/s para moléculas pequeñas. La tensión interfacial de una gota macroscópica de dodecano suspendida en una solución acuosa de 8mM SDS, fue determinada utilizando un tensiómetro de gota rotatoria modelo TGG110-M3 (fabricado por CITEC y el Lab. FIRP en la Universidad de Los Andes, Edo. Mérida).

### 3. RESULTADOS Y DISCUSIÓN

La Fig. 1 muestra la variación del radio cúbico promedio de las emulsiones estudiadas. Es claro que a medida que la fracción de volumen se hace menor, la pendiente decrece sensiblemente. Esto se debe a que el tiempo (y la tasa) de agregación es inversamente proporcional a la densidad numérica de las gotas. A menor número de gotas por unidad de volumen la agregación se ralentiza, y con ella la posible ocurrencia de coalescencia. Si no se forman flóculos suficientemente grandes, el tamaño de las gotas es demasiado pequeño para favorecer la formación de crema durante el tiempo de medición. Aún así, los viales fueron agitados suavemente antes de cada medida de tamaño.

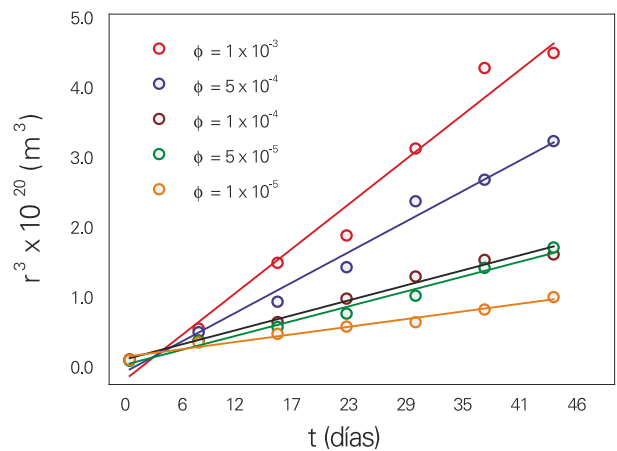

Fig. 1. Variación de r$^3$ vs. t. Las líneas continuas representan el ajuste lineal por mínimos cuadrados.

De acuerdo con lo antes descrito, el proceso de maduración debe convertirse progresivamente en el mecanismo de desestabilización predominante a medida que la fracción de volumen baja. Por tanto la calidad de la regresión lineal debería aumentar (Fig. 1). Curiosamente se encontró que los coeficientes de regresión no variaron de manera monótona (Tabla 1). Esto puede deberse parcialmente a que la frecuencia de las mediciones (una vez por semana) no es lo suficientemente alta.

La Tabla 2 muestra los valores de solubilidad reportados para el dodecano por diferentes autores. Es evidente que sólo el sistema menos concentrado reproduce razonablemente los valores reportados en la bibliografía. Sin embargo no es posible utilizar muestras más diluidas porque se sobrepasa el límite de detección del aparato. Una alternativa es utilizar gotas de mayor tamaño, un surfactante no iónico y un sistema de recirculación que minimice el efecto gravitatorio. No es posible cambiar la composición del aceite a fin de igualar su densidad con la de la fase externa, ya que se modificaría la solubilidad del aceite y la velocidad del fenómeno de maduración debido a la variación de la presión de vapor de la gota. Este efecto ha sido estudiado detalladamente por Kabalnov (2001) y se denomina maduración composicional.

Tabla 1: Valores de solubilidad de dodecano obtenidos a partir de $dr^3/dt$ de cada emulsión.
Los errores de solubilidad fueron calculados por propagación a partir de la Ec. (3).

| ID | φ | dr$^3$/dt (m$^3$/s) | Coeficiente de Regresión | Solubilidad (m$^3$/m$^3$) |
|---|---|---|---|---|
| φ1 | 1 x 10$^{-3}$ | (1,28 ± 0,09) x 10$^{-27}$ | 0.97074 | (3,18 ± 0,22) x 10$^{-8}$ |
| φ2 | 5 x 10$^{-4}$ | (8,74 ± 0,49) x 10$^{-27}$ | 0.98112 | (2,17 ± 0,12) x 10$^{-8}$ |
| φ3 | 1 x 10$^{-4}$ | (4,30 ± 0,23) x 10$^{-27}$ | 0.98283 | (1,07 ± 0,06) x 10$^{-8}$ |
| φ4 | 5 x 10$^{-5}$ | (4,26 ± 0,25) x 10$^{-27}$ | 0.97995 | (1,06 ± 0,06) x 10$^{-8}$ |
| φ5 | 1 x 10$^{-5}$ | (2,19 ± 0,16) x 10$^{-27}$ | 0.97069 | (5,45 ± 0,40) x 10$^{-9}$ |

Tabla 2: Valores reportados de solubilidad para el dodecano.

| Referencia | [McAuliffe, 1969] | [Huibers, 1998] | [Sakai, 2002] | [Kabalnov,1990] | [Franks,1966] |
|---|---|---|---|---|---|
| Solubilidad (en ml/ml) | 5,31 x 10$^{-9}$ | 4,86 x 10$^{-9}$ | 5,46 x 10$^{-9}$ | 5,20 x 10$^{-9}$ | 1,12 x 10$^{-9}$ |

## 4. CONCLUSIÓN

Si bien la teoría LSW permite la implementación de una metodología novedosa para la evaluación de la solubilidad de un líquido hidrofóbico, la Ec. (3) requiere del uso de una fracción de volumen muy reducida (de al menos $\varphi = 10^{-5}$) en el caso de una nanoemulsión aceite/agua.

## 5. BIBLIOGRAFÍA

Frank F, Solute–Water Interactions and the Solubility Behaviour of Long-chain Paraffin Hydrocarbons. *Nature* 210: 87-88 (1966).

Huibers PD, Correlation of the Aqueous Solubility of Hydrocarbons and Halogenated Hydrocarbons with Molecular Structure, *J. Chem. Inf. Comp. Sci.* 38: 283-292 (1998).

Kabalnov AS, Makarov KN, Pertzov AV, Shchukin ED, Ostwald ripening in hydrocarbon emulsions: experimental verification of equation for absolute rates. *J. Colloid Interface Sci.* 138: 98-104 (1990a).

Kabalnov AS, Makarov KN, Shcerbakova OV, Solubility of fluorocarbons in water as a key parameter determining fluorocarbon emulsion stability *J. Fluorine Chem*. 50: 271-284 (1990b).

Kabalnov A, Ostwald ripening and related phenomena, J. Disp. Sci. and Tech. 22: 1-12 (2001).

Lifshitz IM, Slezov V, The kinetics of precipitation from supersaturated solid solution, *J. Phys. Chem. Solids* 19: 35 (1961).

McAuliffe C, Solubility in Water of Normal C9 and C10, Alkane Hydrocarbons. *Science* 163: 478-479 (1969).

Nazarzadeh E, Anthonypillai T, Sajjadi S, On the growth mechanism of nanoemulsions. *J. Colloid Interface Sci.* 397: 154-162 (2012).

Rahn-Chique K, Puertas AM, Romero-Cano MS, Rojas C, Urbina-Villalba G, Nanoemulsion stability: experimental evaluation of the flocculation rate from turbidity measurements. *Adv. Colloids and Interface Sci.* 178: 1-20 (2012).

Sakai A, Kamogawa K, Nishiyama K, Sakai H y Abe M, Molecular diffusion of oil/water emulsions in surfactant-free conditions. *Langmuir* 18: 1985-1990 (2002).

Taylor P. Ostwald ripening in emulsions. *Adv. Colloids and Interface Sci.* 75: 107-163 (1998).

Thomson W, On the equilibrium of vapour at a curved surface of liquid. *Philosophical Magazine*, Series 4, 42: 448-452 (1871).

Urbina-Villalba, G. El fenómeno de maduración de Ostwald. Predicciones de las simulaciones sobre la evolución del radio cúbico promedio de una dispersión aceite/agua. *RCEIF* 3: 1-21 (2014).

Wagner C, Theorie der Alterung von Niedershlagen durch Umlosen, *Z. Elektrochemie* 65: 581-591 (1961).

Weers JG. Molecular diffusion in emulsions and emulsion mixtures. En: Modern Aspects of Emulsions Science. Cambridge, BP Binks (Ed.), Capítulo 9. *The Royal Society of Chemistry Publications* (1999).